\documentclass[useAMS,english,a4paper,12pt]{article}
\usepackage{amssymb,amsmath}
\usepackage{xfrac}
\usepackage{graphicx}
\usepackage{subfigure}
\usepackage{float}
\usepackage{xcolor}
\usepackage{dcolumn}
\usepackage{bm}
\usepackage{dblfloatfix}
\usepackage{fixltx2e}
\usepackage{tikz}
\usetikzlibrary{arrows}
\usepackage{epsfig}
\usepackage{enumitem}
\usepackage{ulem}
\usepackage[utf8]{inputenc}
\usepackage[english]{babel} 
\usepackage{svg}
\usepackage[sorting=none,citestyle=phys,minbibnames=1,maxbibnames=10,backend=biber]{biblatex}
\DeclareFieldFormat[article,periodical]{volume}{\mkbibbold{#1}} 
\usepackage[babel]{csquotes}
\usepackage[mode=buildnew]{standalone}
\usepackage{import,pgfplots}
\usetikzlibrary{spy}
\usepgfplotslibrary{groupplots}
\usepackage{pdfpages}
\usepackage{float}
\usepackage{soul} 
\usepackage{authblk}
\pgfplotsset{compat=1.17} 
\usepackage{siunitx}
\usepackage{lineno} 

\title{Spatial profile of plasma temperature generated by discharge in 3D printed capillary}
\author[1]{Niv Barkai}
\author[1]{Noa kliss}
\author[2]{Yair Ferber}
\author[1]{Yoav Raz}
\author[1]{Rotem Liran}
\author[1]{Mordechai Botton}
\author[1] {Arie Zigler}
\date{September 2023}
\affil[1]{Racah Institute of Physics, Hebrew University, 91904 Jerusalem, Israel}
\affil[2]{National Laser Facility (NLF), Soreq NRC, Yavne 81800, Israel}
\usepackage[draft]{changes} 

\begin{document}

\maketitle
\begin{abstract}
A method for evaluating the radial temperature profile of a plasma channel is presented. 
The radiation spectrum of mixture of Hydrogen Nitrogen plasma in capillary discharge is collected. Assuming a local thermal equilibrium, the temperature is derived 
from the ratio of two emission lines ($N^+$,$N^{2+}$).
The method presented here does not rely on Stark broadening of the emitted $H_\alpha$ 
line hence can be utilized in a wide range of plasma constituents and densities, leading to an  
improved control of the plasma channel parameters as required by future laser wakefield acceleration schemes.
\end{abstract}

\pagebreak
\section{Introduction}
Plasma channels generated in capillaries discharges are considered to be
the backbone in promising next generation high-energy particle accelerators \parencite{TajimaMalka2020}. 
In the laser wakefield acceleration (LWFA) process \parencite{TajimaDawson1979,Tajima2020}, the laser pulse is focused onto a gas or plasma target, 
creating within the plasma a highly non-uniform electron density moving closely behind the laser pulse. 
This density non-uniformity is accompanied by a high-intensity localized electric field in the plasma, 
known as a wakefield. Injecting electron bunches synchronized with the wakefield the former can be 
trapped by the latter and hence be accelerated by it to significantly higher energy. 
Record level of multi-GeV electron bunches using LWFA was demonstrated in various experiments \parencite{Gonsalves2019,Leemans2014,Kim2013}.
Underlying all these Innovative technologies is the requirement that the laser pulse propagates
in the interaction region for lengths many times longer than Rayleigh lengths of 
the focused laser pulse. Such is the case in plasma channels formed in gas-filled capillaries. 
Nonetheless, a particular transverse profile of the plasma channel, more specifically, a minimum 
value on the axis and increasing towards the wall is required \parencite{Bobrova2013}. 
Accordingly, an extensive effort was and still is being given to the diagnostics of the
plasma channel parameters.
One prominent method relies on the spectroscopic properties of the plasma,
more specifically on the Stark broadening of the emitted $H_\alpha$ 
line \parencite{Griem2,Morgan1994,Oh2010}.
Other methods used the interferometric approach, in one or more 
“colors”  \parencite{Spence2000,Jones2003}.
Other methods utilized the dependence of the group velocity of a propagating electromagnetic pulse 
in the plasma channel on the local density \parencite{Daniels2015,Tilborg2018}. 
 Combinations of the above-mentioned methods were employed as well \parencite{Jang2011,Garland2021}. 
Spectroscopy-based measurements albeit detailed are 
correlated to Hydrogen or Hydrogen like cases. Furthermore, direct 
measurement of density is limited to values above $10^{17}cm^{-3}$. 
As the requirements of the plasma channel may vary in parameters (contents and density), 
a more robust diagnostics method may prove useful. 

In this paper, we present a utilization of such a method where the
temperature profile of the plasma channel is measured.
The method presented here is facilitated by  the conditions of local thermal equilibrium (LTE).
The temperature measurement approach is based on comparing the 
intensities of two separate emission lines. These lines can be transition in two sequential ionization stages by different ionization levels, or alternatively by two different excitation of the same ion. 
Thus it can be applied to the broad range of gases in capillary discharge and need not include Hydrogen.

Subsequently, the density can inferred by relying on the reasonable assumption of local thermal 
equilibrium, which guarantees a unique relation between the temperature 
and density by means of the equation of state (that more often than not is
of the ideal gas).
For the purpose of demonstrating this method, we present spatial and time-dependent measurements 
of the plasma temperature. accompanied by direct measurement of the density using the conventional Stark broadening.

\section{Experimental setup}
\begin{figure}[H]
    \centering
    \includegraphics[width=1.4\textwidth]{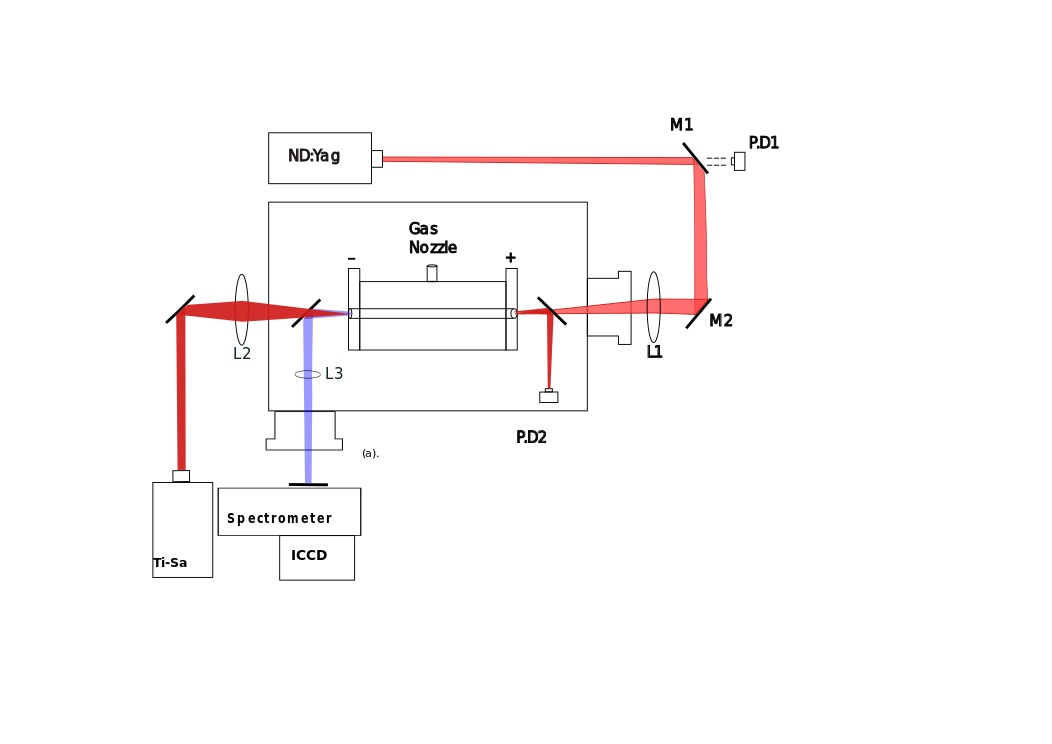}
        \caption{The entire system scheme is comprised by parts :
        The ND:Yag laser is used for the triggering scheme with the high voltage generator, The Ti-Sa is used as a time 
        measurement of the channel using the P.D2 photodiode. The Spectrometer collects the blue light from the discharge using L3 lens.}
    \label{fig:scheme_module_exper_1}
\end{figure}

The experiments were carried out at the High Intensity Laser Lab at the Hebrew University of Jerusalem. 
The experimental setup is depicted in figure \ref{fig:scheme_module_exper_1}. 
A 3D-printed PolyGet capillary is positioned inside a vacuum chamber and a controlled flow of Hydrogen-Nitrogen gases was flowing through it. The capillary size is of [length $\SI{5}{cm}$,  diameter $\SI{500}{\mu m}$].
Discharge in the gas-filled capillary is initiated by a laser pulse injected into the capillary, 
followed by a high-voltage breakdown along the capillary that produces the actual plasma channel.

The triggering laser is a Q-switched Nd:Yag laser (\SI{30}{\mJ}, \SI{1064}{\nm}, \SI{10}{\ns}) and is 
directed by a two-mirror setup into the capillary entrance (figure\ref{fig:scheme_module_exper_1}). 
Part of its energy is used as a triggering signal to both the high-voltage circuit and the measuring system. 
The discharge gas consists of a nitrogen-hydrogen mixture (95-5 percent) at a pressure of 1.5 $bar$, with gas flow maintained for 35 $ms$. 
Synchronously following this trigger[Nd:Yag] is a high-voltage pulse (up to \SI{15}{kV}, \SI{0.6}{\us}) 
applied on the capillary between two electrodes mounted at its ends. The high voltage pulse generates a breakdown in the plasma and hence   
the creation of a plasma discharge in the capillary, which effectively closes an electrical circuit \parencite{Palchan2007}. 
Prior to the temperature measurements and in order to establish the proper time delay gating, a pulse 
train from a Ti-Sa laser (84MHz 800nm) was injected into the generated channel (see Fig. 1). The 
output intensity was collected and measured.
During the propagation of the laser inside the capillary, some part of the light is lost to the walls, when conditions of the plasma channel are met there is internal reflection inside the capillary and more light is being measured with P.D2 

The radiation emitted from the plasma discharge was measured using a spectrometer (Acton Research Corp 
spectrometer coupled to Andor ICCD camera).
The discharge and the spectrometer  are synchronized to collect the radiation by a Stanford gating device. 
The spectrometer is tuned to the specific lines of the plasma involved here, The $N^{2+}$ and $N^{+}$ ions. 
In order to establish a direct connection to the conventional methods, the average density was measured by Stark-broadening 
of the $H_{\alpha}$ line.
As will be shown later an exact knowledge of the averaged density is not required in our method.
The spectrometer's gating time was 50 to 20 $ns$ by width and the time delay was interchanged regarding the guiding measurements.  
The temperature profile measurement scheme utilizes the Acton SP-300i spectrometer and Andor new-iStar ICCD camera.

\section{Results and discussion}
\begin{figure}[H]
    \centering
    \includegraphics[width=\columnwidth]{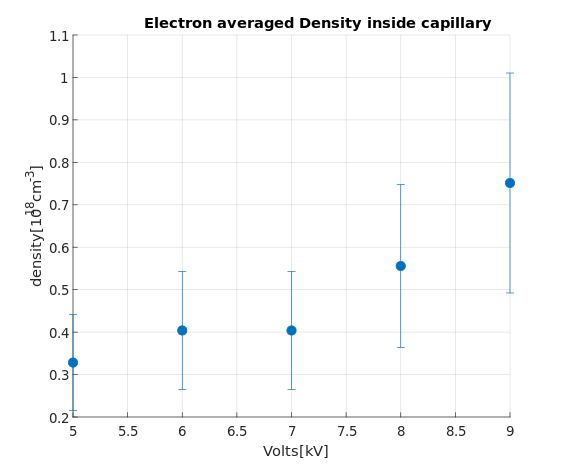}
    \caption{Average electron density as a function of breakdown Voltage.}
    \label{fig:Density}
\end{figure}

In order to relate the method presented here to previously published method, we begin by presenting the 
electron density measurements carried out by the conventional Stark-broadening of 
the H\textsubscript{$\alpha$} Balmer series (the wavelength is \SI{656.26}{\nm}).
The derived average density as a function of the high voltage is shown in Fig. \ref{fig:Density}. 
These values can be used in the analysis of the temperature measurements as will be demonstrated shortly.
Temperature measurements which are based on the intensity ratio of two separate emission lines require 
that the plasma would be in a local thermal equilibrium (LTE) or near-LTE, and that the radiation loss rate would 
be small compared to the collision rate \parencite{Griem}. 
Both conditions are met in this case when the measurements take place sometime after the initiation of 
the plasma discharge when the ionization fraction is high (close to one) and the primary heating process 
is Ohmic \parencite{Steinhauer2006,Bobrova2002}. 
Accordingly, emission line measurements were done after the discharge current 
reached its peak value. Out of the multitude of possible emission lines we chose to concentrate our 
attention on two which are expected to be dominant given the expected temperature scale ($\lesssim 10eV$). 
The lines of two distinct ionization levels of Nitrogen, namely $N^{+}$ and $N^{2+}$ are 
relatively close in wavelength so the spectrometer can be tuned to collect both of them (see Table 1 for details). 

\begin{table}[h!]
\centering
\begin{tabular}{||c c c c c||} 
\hline
 wavelength[$\lambda$] & $f_{ik}$ & $g_kA_{ki}$ & $^JL_S$ &Ion\\
 \hline\hline
 571.077nm & $5.74e{-2}$ & 5.85$e7$ & $^3D\rightarrow\ ^3P^0$ & $N^{+}$\\ 
 581.779nm & 8.94$e{-3}$ & 7.04$e6$ & $^2D\rightarrow ^2P^0 $& $N^{2+}$\\
 
\hline
\end{tabular}
\caption{The parameters of the two measured lines $N^{2+}$ and $N^+$ \parencite{NIST}.
$f_{ik}$ is the oscillator strength which corresponds to the probability for the transition in certain ion, the parameter $g_kA_{ki}$ is the degeneracy and Einstein coefficient which relates to spontaneous emission, and the $^JL_S$ is the spectroscopic symbol which gives the total angular momentum of the certain state $J=L+S$  
}
\label{table:1}
\end{table}

A typical result of the radiation emitted in these lines is shown in 
Fig. \ref{lines of $ N^+$ and $N^{2+}$} for two values of breakdown voltage. From these figures, it is 
clearly evident that the emission intensity of the $N^+$ is approximately unchanged while the emission 
intensity of the $N^{2+}$ significantly increased when the breakdown voltage is changed from 7kV to 9kV

\begin{figure}[H]
    \centering
    \includegraphics[width=\columnwidth]{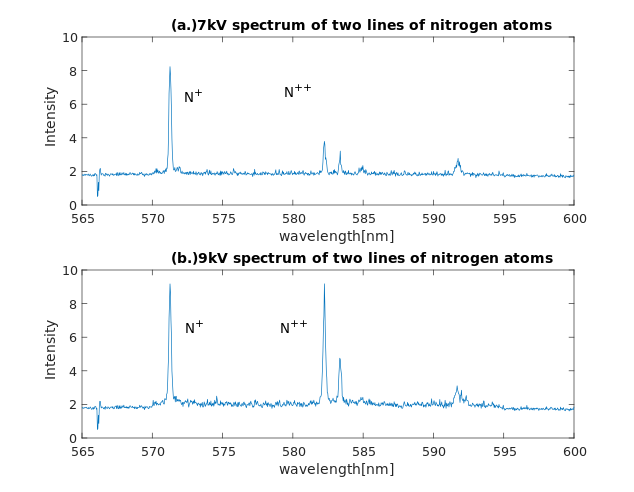}
    \caption{
    (a.)7kV spectrum of different ionization degrees of nitrogen\newline
    (b.) 9kV spectrum of different ionization degrees of nitrogen
    }
    \label{lines of $ N^+$ and $N^{2+}$}
\end{figure}

\begin{figure}[H]
    \centering
    \includegraphics[width=\columnwidth]{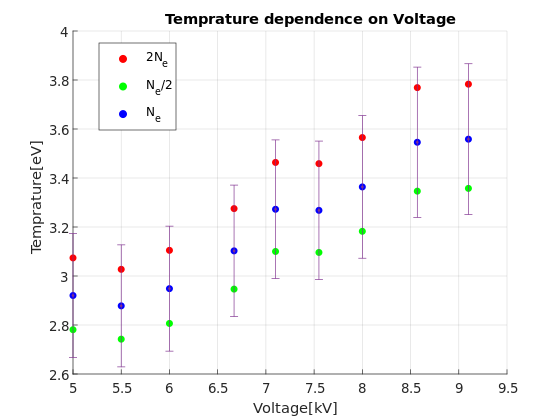}
    \caption{Temperature dependence on Voltage in our capillary, each point is calculated with the appropriated density using figure \ref{fig:Density}, The multiple points correspond to changing the density.
    One can see that only the scaling changes not the trend of the points.}
    \label{fig:Temp(V)}
\end{figure}

\begin{figure}[H]
    \centering
    \includegraphics[width=\columnwidth]{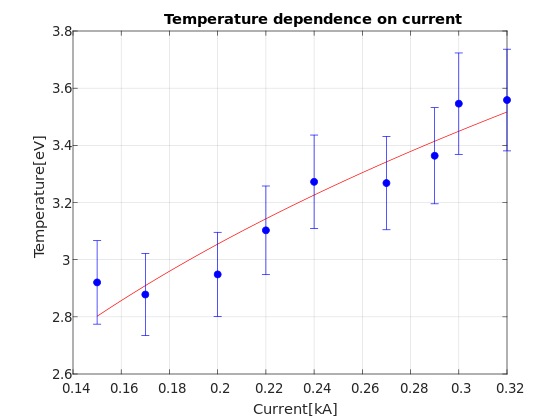}
    \caption{Temperature dependence on the current with a theoretical fit \parencite{Bobrova2002}}
    \label{T(I)}
\end{figure}

Relating the emission intensity from a line to its occupation fraction, and utilizing the well know 
Saha-Boltzmann equation,\parencite{Griem2} we obtain the following expression for the intensity ratio $R$:
\begin{equation}
    R=\frac{i'}{i}=\frac{w'A'g'}{\sqrt{\pi} wAg}(4\pi a_{0}^3 N_e)^{-1}(\frac{k_BT}{E_H})^{3/2} \times exp(-\frac{E'+E_{\infty}-E-\Delta E_{\infty}}{k_BT})
\end{equation}
Here $\omega,\omega'$ are the frequencies of the two levels (related to $E,E'$ for each ion respectively),  $A, A'$ are the Einstein coefficients, $g,g'$ are the degeneracy levels, $a_0$ is the Bohr radius $E_H$ is ionization energy of hydrogen $E_{\infty}$ is the difference in ionization energy between the certain ions,  $\Delta E_\infty$ is the reduction to the ionization energy due to charge loss, and $N_e$ is the electron density.
For the calculations of the temperature from the intensity ratio measured for each value of the breakdown voltage 
we use the averaged electron densities, $N_e$, presented in Fig. \ref{fig:Temp(V)}. 
The measurements are being averaged spatially by the ICCD, and the gating time is 50ns, 
so there is an average over space and over time. 
Noting that the dependence  of the above ratio on the density is linear while on the temperature it 
is exponential, we expect that the temperature values will only be slightly affected by systematic errors 
in the density values. This effect is demonstrated in Fig. 4, where the temperature is calculated with the 
lowest and highest values of the electrons density. The exact values of the derived temperature slightly 
change, however, the tendency of the graph remains the same. 
Following \parencite{Griem2}, the error is estimated to be 10 percent.
Following previous work the current-voltage relation is expected to be given by Eq. 39 in \parencite{Bobrova2002}
\begin{equation}
    T(0)\approx 5.7{\left( \frac{I[kA]}{R_0[mm]}^{2/5} \right)} eV
\end{equation}
Here $R_0$ is the radius of the capillary.
A plot of the inferred temperature as a function of the measured current is presented in \ref{T(I)} and compared to the theoretical fit. 
The correspondence is very good.

Next we address the spatial profile of the temperature. The spectrum is imaged to a slit in the spectrometer entrance, 
and two images of the capillary cross section are obtained for the two lines.
The ICCD is not averaging with respect to space, accordingly, each pixel correspond to a pixel of the cross section of the capillary.
Relating two corresponding images of the cross section of the capillary and obtaining ratio of lines per pixel 
we got the ratio with respect to the location of the cross section of the capillary.
Following that, the temperature spatial profile is obtained by repeating the ratio calculations for the same lines for each point. 
The results of these calculations are presented in Fig. \ref{fig:temperature_radial_profile}.
In order to look for the radial temperature profile we needed to reduce the gating time from 50ns to  20ns.
To achieve that, a detailed knowledge of the time profile of the channel is required. For that purpose, we employed a Ti:Si laser (84MHz-10ns) focused into the capillary. A photodiode collected the laser pulse at the capillary exit.
\begin{figure}[H]
    \centering
    \includegraphics[width=\columnwidth]{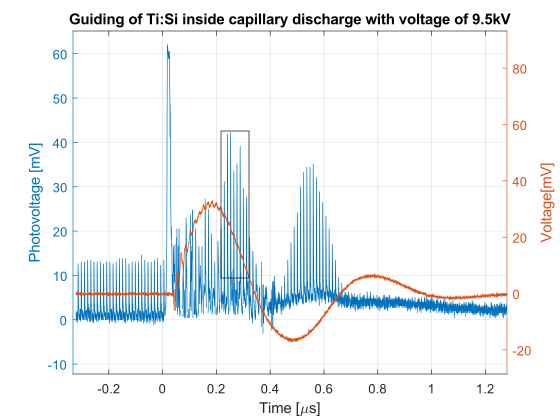}
    \caption{guiding of the oscillator inside our capillary the orange represents the current discharge, and the blue is the local oscillator being amplified and guided inside our capillary}
    \label{fig:guiding oscillator}
\end{figure}

Before the discharge, one can look in figure \ref{fig:guiding oscillator} and spot the Ti:Si with laser pulse every 10$ns$ 
(voltage pulses in time before the discharge, $t<0$ in Fig. 6). Following the discharge, two distinct pulse trains with higher voltage 
are observed (around $0.3\mu s$ and $0.5\mu s$). As the laser energy collected by the photodiode is unchanged, the increase
in the voltage output is related to a corresponding increase of the collected intensity.
Accordingly, the laser light is being guided inside the capillary, so there is no loss to the walls of the capillary,  which is the result of a higher output in the photodiode.
This measurement gives us the required time-dependent data of the lifetime of the channel with a resolution of 10$ns$.
We observed two gains after the ignition of plasma. The first one after $\simeq$230ns, was used for the temperature profile measurements.
A second channel after $\simeq$500ns, 
was observed only for higher breakdown voltages. It is probably related to the 
longer lifetime of the channel in higher breakdown voltages. As it did not appear in all voltages we did not use it for
the temperature profile.

The radial profile of the electron density was measured in \parencite{Raz2021} demonstrating the conjugate profile of the temperature profile presented here. The multiplication of the two shows that the pressure is almost constant across the capillary cross-section.

\begin{figure}[H]
    \centering
    
    \includegraphics[width=\columnwidth]{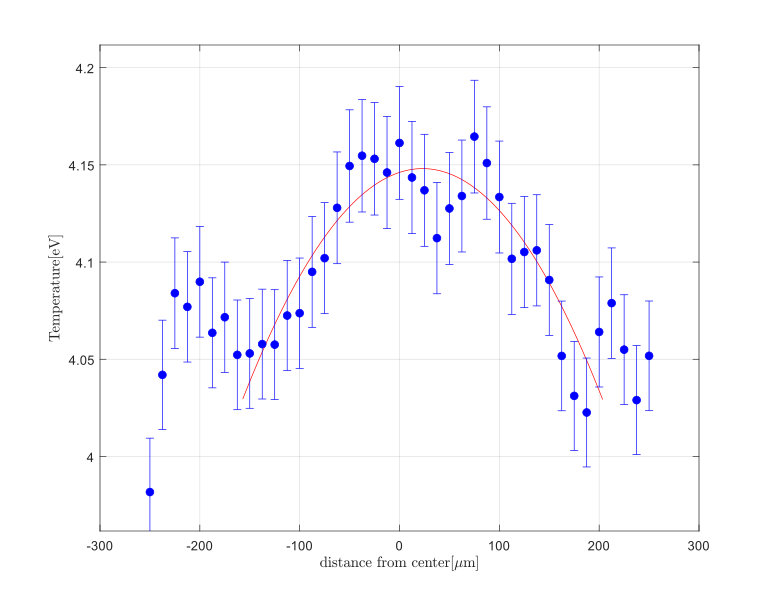}
    
    \caption{Temperature profile measurements, measured with a gating of 20ns at 280ns delay from the Nd-Yag laser starting the breakdown.}
    \label{fig:temperature_radial_profile}
\end{figure}

Viewing figure \ref{fig:temperature_radial_profile} one can spot the almost parabolic appearance of the temperature profile 
in the radii between values of $-50:50\mu m$. This profile represents the core of the plasma channel. For larger radii values 
the profile changes as we move closer to capillary walls.

\section{Conclusion} 
In conclusion, the spectroscopic measurements of capillary discharge were presented. Using a 3D-printed 
capillary filled with nitrogen/hydrogen gas at a pressure of $1*10^{-5}$ Torr. The results showed that 
the emission from the plasma was due to excited atomic states of nitrogen, and an excitation 
temperature of around $3.6eV$ was calculated and correlated to theory with \parencite{Bobrova2002}.
This method provided insight into the plasma characteristics within the discharge and was found to be 
a useful way of determining the existence of a plasma channel within the capillary. 
The spatial profile of the plasma lens was obtained and demonstrated. The method is expected to be useful for any NLTE LTE plasma.
Future work may involve improving the lifetime of the discharge by using stronger capacitors and 
investigating the physical mechanism of the appearance of the second channel we observed, 
possibly improving the model for capillary discharge.

\printbibliography

@article{TajimaMalka2020,
    title = {Laser plasma accelerators},
    author = {T Tajima and V Malka},
    journal = {Plasma Phys. Control. Fusion},
    doi = {10.1088/1361-6587/ab6da4},
    url = {https://dx.doi.org/10.1088/1361-6587/ab6da4},
    year = {2020},
    month = {feb},
    publisher = {IOP Publishing},
    volume = {62},
    number = {3},
    pages = {034004},
}

@article{TajimaDawson1979,
  title = {Laser Electron Accelerator},
  author = {Tajima T. and Dawson J. M.},
  journal = {Phys. Rev. Lett.},
  volume = {43},
  issue = {4},
  pages = {267--270},
  numpages = {0},
  year = {1979},
  month = {Jul},
  publisher = {American Physical Society},
  doi = {10.1103/PhysRevLett.43.267},
  url = {https://link.aps.org/doi/10.1103/PhysRevLett.43.267}
}

@article{Tajima2020,
	author = {Tajima, T. and Yan, X. Q. and Ebisuzaki, T.},
	date = {2020/05/06},
	doi = {10.1007/s41614-020-0043-z},
	id = {Tajima2020},
	isbn = {2367-3192},
	journal = {Reviews of Modern Plasma Physics},
	number = {1},
	pages = {7},
	title = {Wakefield acceleration},
	url = {https://doi.org/10.1007/s41614-020-0043-z},
	volume = {4},
	year = {2020}}

@article{Kim2013,
  title = {Enhancement of Electron Energy to the Multi-GeV Regime by a Dual-Stage Laser-Wakefield Accelerator Pumped by Petawatt Laser Pulses},
  author = {Kim, Hyung Taek and Pae, Ki Hong and Cha, Hyuk Jin and Kim, I Jong and Yu, Tae Jun and Sung, Jae Hee and Lee, Seong Ku and Jeong, Tae Moon and Lee, Jongmin},
  journal = {Phys. Rev. Lett.},
  volume = {111},
  issue = {16},
  pages = {165002},
  numpages = {5},
  year = {2013},
  month = {Oct},
  publisher = {American Physical Society},
  doi = {10.1103/PhysRevLett.111.165002},
  url = {https://link.aps.org/doi/10.1103/PhysRevLett.111.165002}
}

@article{Morgan1994,
  title = {Spectroscopic measurements of electron density and temperature in polyacetal-capillary-discharge plasmas},
  author = {Morgan, C. A. and Griem, H. R. and Elton, R. C.},
  journal = {Phys. Rev. E},
  volume = {49},
  issue = {3},
  pages = {2282--2290},
  numpages = {0},
  year = {1994},
  month = {Mar},
  publisher = {American Physical Society},
  doi = {10.1103/PhysRevE.49.2282},
  url = {https://link.aps.org/doi/10.1103/PhysRevE.49.2282}
}

@article{Oh2010,
author = {Oh,Seong Y.  and Uhm,Han S.  and Kang,Hoonsoo  and Lee,In W.  and Suk,Hyyong },
title = {Temporal evolution of electron density and temperature in capillary discharge plasmas},
journal = {Journal of Applied Physics},
volume = {107},
number = {10},
pages = {103309},
year = {2010},
doi = {10.1063/1.3415553},
URL = { 
        https://doi.org/10.1063/1.3415553
    
},
eprint = { 
        https://doi.org/10.1063/1.3415553
}
}

@article{Spence2000,
  title = {Investigation of a hydrogen plasma waveguide},
  author = {Spence, D. J. and Hooker, S. M.},
  journal = {Phys. Rev. E},
  volume = {63},
  issue = {1},
  pages = {015401},
  numpages = {4},
  year = {2000},
  month = {Dec},
  publisher = {American Physical Society},
  doi = {10.1103/PhysRevE.63.015401},
  url = {https://link.aps.org/doi/10.1103/PhysRevE.63.015401}
}

@article{Jones2003,
author = {Jones,T. G.  and Ting,A.  and Kaganovich,D.  and Moore,C. I.  and Sprangle,P. },
title = {Spatially resolved interferometric measurement of a discharge capillary plasma channel},
journal = {Physics of Plasmas},
volume = {10},
number = {11},
pages = {4504-4512},
year = {2003},
doi = {10.1063/1.1615578},
URL = { 
        https://doi.org/10.1063/1.1615578
},
eprint = { 
        https://doi.org/10.1063/1.1615578
}
}

@article{Garland2021,
author = {Garland,J. M.  and Tauscher,G.  and Bohlen,S.  and Boyle,G. J.  and D’Arcy,R.  and Goldberg,L.  and Põder,K.  and Schaper,L.  and Schmidt,B.  and Osterhoff,J. },
title = {Combining laser interferometry and plasma spectroscopy for spatially resolved high-sensitivity plasma density measurements in discharge capillaries},
journal = {Review of Scientific Instruments},
volume = {92},
number = {1},
pages = {013505},
year = {2021},
doi = {10.1063/5.0021117},
URL = { 
        https://doi.org/10.1063/5.0021117
},
eprint = { 
        https://doi.org/10.1063/5.0021117
}
}

@article{Jang2011,
author = {Jang,D. G.  and Kim,M. S.  and Nam,I. H.  and Uhm,H. S.  and Suk,H. },
title = {Density evolution measurement of hydrogen plasma in capillary discharge by spectroscopy and interferometry methods},
journal = {Applied Physics Letters},
volume = {99},
number = {14},
pages = {141502},
year = {2011},
doi = {10.1063/1.3643134},
URL = { 
        https://doi.org/10.1063/1.3643134
},
eprint = { 
        https://doi.org/10.1063/1.3643134
}
}

@article{Daniels2015,
author = {Daniels,J.  and van Tilborg,J.  and Gonsalves,A. J.  and Schroeder,C. B.  and Benedetti,C.  and Esarey,E.  and Leemans,W. P. },
title = {Plasma density diagnostic for capillary-discharge based plasma channels},
journal = {Physics of Plasmas},
volume = {22},
number = {7},
pages = {073112},
year = {2015},
doi = {10.1063/1.4926825},
URL = { 
        https://doi.org/10.1063/1.4926825
},
eprint = { 
        https://doi.org/10.1063/1.4926825
}
}

@article{Tilborg2018,
title = {Density characterization of discharged gas-filled capillaries through common-path two-color spectral-domain interferometry},
author = {van Tilborg, J. and Gonsalves, A. J. and Esarey, E. H. and Schroeder, C. B. and Leemans, W. P.},
abstractNote = {},
doi = {10.1364/OL.43.002776},
journal = {Optics Letters},
number = 12,
volume = 43,
place = {United States},
year = {2018},
month = {6}
}

@article{Steinhauer2006,
   author = {L. C. Steinhauer,W. D. Kimura},
   doi = {10.1103/PhysRevSTAB.9.081301 },
   issn = {081301},
   issue = {9},
   journal = {Phys. Rev. ST Accel. Beams},
   month = {2},
   title = {Quasistatic capillary discharge plasma model},
   volume = {9},    
   year = {2006},
}

@article{Gonsalves2019,
   author = {A. J. Gonsalves and K. Nakamura and J. Daniels and C. Benedetti and C. Pieronek and T. C. H. de Raadt and S. Steinke and J. H. Bin and S. S. Bulanov and J. van Tilborg and C. G. R. Geddes and C. B. Schroeder and Cs. Tóth and E. Esarey and K. Swanson and L. Fan-Chiang and G. Bagdasarov and N. Bobrova and V. Gasilov and G. Korn and P. Sasorov and W. P. Leemans},
   doi = {10.1103/PhysRevLett.122.084801},
   issn = {0031-9007},
   issue = {8},
   journal = {Physical Review Letters},
   month = {2},
   title = {Petawatt Laser Guiding and Electron Beam Acceleration to 8 GeV in a Laser-Heated Capillary Discharge Waveguide},
   volume = {122},
   year = {2019},
}

@article{Bobrova2013,
   author = {N. A. Bobrova and P. V. Sasorov and C. Benedetti and S. S. Bulanov and C. G. R. Geddes and C. B. Schroeder and E. Esarey and W. P. Leemans},
   doi = {10.1063/1.4793447},
   issn = {1070-664X},
   issue = {2},
   journal = {Physics of Plasmas},
   month = {2},
   title = {Laser-heater assisted plasma channel formation in capillary discharge waveguides},
   volume = {20},
   year = {2013},
}

@article{Palchan2007,
   author = {T. Palchan and D. Kaganovich and P. Sasorov and P. Sprangle and C. Ting and A. Zigler},
   doi = {10.1063/1.2472525},
   issn = {00036951},
   issue = {6},
   journal = {Applied Physics Letters},
   title = {Electron density in low density capillary plasma channel},
   volume = {90},
   year = {2007},
}

@article{Bobrova2002,
   author = {N.A. Bobrova and A.A. Esaulov and J.-I. Sakai and P.V. Sasorov and D.J. Spence and A. Butler and S.M. Hooker and S.V. Bulanov},
   doi = {10.1103/PhysRevE.65.016407},
   issue = {1},
   journal = {Phys. Rev. E},
   title = {Simulations of a hydrogen-filled capillary discharge waveguide},
   volume = {65},
   year = {2002},
}

@article{Leemans2014,
   author = {W. P. Leemans and A. J. Gonsalves and H.-S. Mao and K Nakamura and C Benedetti and C. B. Schroeder and Cs. Tóth and J Daniels and D. E. Mittelberger and S. S. Bulanov and J.-L. Vay and C. G. R. Geddes and E Esarey},
   doi = {10.1103/PhysRevLett.113.245002},
   issue = {24},
   journal = {Physical Review Letters},
   month = {12},
   pages = {245002},
   publisher = {American Physical Society},
   title = {Multi-GeV Electron Beams from Capillary-Discharge-Guided Subpetawatt Laser Pulses in the Self-Trapping Regime},
   volume = {113},
   url = {https://link.aps.org/doi/10.1103/PhysRevLett.113.245002},
   year = {2014},
}

@article{Raz2021,
   author = {Yoav Raz and Ehud Behar and Yair Ferber and  Angelo Biagioni and Mario Galletti and Maria Pia Anania and Riccardo Pompili and Costa Gemma2 and Arie Zigler},
   doi = {10.1088/PhysRevLett.113.245002},
   issue = {24},
   journal = {Plasma Research Express},
   month = {6},
   publisher = {IOP},
   title = {Low jitter parabolic profile low density plasma channel in 3D printed
gas filled capillary},
   volume = {113},
   url = {https://iopscience.iop.org/article/10.1088/2516-1067/ac0d4b},
   year = {2021},
}

@book{Griem,
    year = {1997},
    publisher = {Cambridge},
    title = {Principles of Plasma Spectroscopy},
    author = {Hans R. Griem},
    isbn = {0521455049},
    series = {Cambridge monographs on plasma physics},
}

@book{Griem2,
    year = {1974},
    publisher = {Cambridge},
    title = {Spectral Line Broadening
by Plasmas},
    author = {Hans R. Griem},
    isbn = {0-12-302850- 7},
    series = {Cambridge monographs on plasma physics},
}

@article{NIST,
    title={NIST Atomic Spectra Database},
    url= {e http://physics.nist.gov/PhysRefData/ASD/index.html},
}

\end{document}